\documentclass[fleqn,usenatbib,useAMS]{mnras}

\usepackage{graphicx}	
\graphicspath{ {./images/} }
\usepackage{amsmath}	
\usepackage{amssymb}	
\usepackage{multicol}        
\usepackage{bm}		
\usepackage{pdflscape}	
\usepackage{float}
\usepackage{dblfloatfix}
\usepackage{ulem}
\usepackage[utf8]{inputenc} 
\usepackage[T1]{fontenc}

\usepackage[T1]{fontenc}
\usepackage{ae,aecompl}

\usepackage{comment}
\usepackage[ruled,vlined]{algorithm2e}

\usepackage{newtxtext,newtxmath}

\title[Detecting and analysing the topology of the cosmic web]{Detecting and analysing the topology of the cosmic web with spatial clustering algorithms I: Methods}

\author[Dimitrios Kelesis et al.]{\parbox{\linewidth}{
Dimitrios Kelesis$^{1,5}$\footnotemark[1],
Spyros Basilakos$^{2,3,4}$,
Vicky Papadopoulou Lesta$^{4}$,
Dimitris Fotakis$^{1}$\footnotemark[1]
and Andreas Efstathiou$^{4}$}
\\
\\
$^{1}$School of Electrical and Computer Engineering, National Technical University of Athens, Athens, 15780, Greece  \thanks{ \{{\sf fotakis, kelesis\}@cs.ntua.gr}}\\
$^{2}$National Observatory of Athens, Lofos Nymfon, 11852 Athens, Greece\\
$^{3}$Academy of Athens, Research Center for Astronomy and Applied Mathematics, Soranou Efesiou 4, 11527, Athens, Greece\\
$^{4}$School of Sciences, European University Cyprus, Diogenes Street, Engomi, 1516 Nicosia, Cyprus\\
$^{5}$National Center for Scientific Research ``Demokritos"
}
\date{\today}

\pubyear{2021}

\begin{document}
\label{firstpage}
\pagerange{\pageref{firstpage}--\pageref{lastpage}}
\maketitle

\begin{abstract}
\normalem
In this paper we explore the use of spatial clustering algorithms as a new computational approach for modeling the cosmic web. We demonstrate that such algorithms are efficient in terms of computing time needed.
We explore  three distinct spatial methods which we suitably adjust for (\emph{i}) detecting the topology of the cosmic web and (\emph{ii}) categorizing various cosmic structures as voids, walls, clusters and superclusters based on a variety of topological and physical criteria such as the physical distance between objects, their masses  and local densities.
The methods explored are (1)
   a new spatial method called  \textit{Gravity Lattice}\,;
  (2)   a modified version of  another spatial clustering algorithm, the ABACUS;
  and (3) the well  known spatial clustering algorithm HDBSCAN.
   We utilize HDBSCAN in order to detect cosmic structures and categorize them using their overdensity. We demonstrate that the ABACUS method can be combined with the classic DTFE method  to obtain  similar results in terms of the achieved accuracy with about an order of magnitude less computation time. To further solidify our claims, we draw insights from the computer science domain and compare the quality of the results with and without the application of our method. Finally, we further extend our experiments and verify their effectiveness by showing their ability to scale well with different cosmic web structures that formed at different redshifts.
\end{abstract}



\section{Introduction}
\footnotetext[1]{Given a set of points located in space, spatial clustering algorithms \citep{miller2009geographic} partition them into groups of closely placed points of high density.}

The combination of observational data from redshift surveys of galaxies \citep{Colless_2001,Tegmark_2004,Driver_2009,Guzzo_2014} with high quality $\Lambda$CDM simulations
has led to great progress in understanding the topological structure and connectivity of the cosmic web
(cf. \citealt{Bond_1996,Einasto_1997,vandeWeygaert2008}). In brief, matter which includes galaxies, intergalactic gas, and dark matter in the universe is not randomly distributed, but tends to form galaxy clusters, filaments and walls which are surrounded by low density regions, the so called voids (c.f. \citealt{Colberg_2008,Weygaert_2011,Nevenzeel_2013,Cautun_Weygaert_2014,Pranav_2017,Feldbrugge_2019,Pranav_2019,Wilding_2021} and references therein). In this context, galaxy clusters are the most massive virialized systems known having the characteristic feature to aggregate in larger groups, the so-called filaments or superclusters of galaxies.

Filaments occupy a special position in
the hierarchy of structure formation since they are
the largest mass units we observe today.
The fact that they have been seeded by density
perturbations of the largest scale ($\sim 100h^{-1}$Mpc)
make them ideal structures with which we can study
the features of the matter fluctuations that gave rise to them.
On the other hand, it has been found that voids
are the most common features of the large scale
structure of the universe due to the fact that they
cover more than half of its volume. Only in the last two decades we 
have galaxy surveys which are large enough to allow
systematic investigations of voids and the galaxies inside them.
For a recent review on various aspects of cosmic networks we refer the 
reader to the paper by \citet{Weygaert_2014} and the references therein.

A large number of computational methods
have been proposed in the literature
which aim to aid the understanding of
the topological and geometrical pattern of the universe as a
whole \citep{Zeldovich_1982,Dekel_1985,Klypin_1993,Sahni_1997,Sahni_1998,Basilakos_Plionis_Rowan-Robinson_2001,Basilakos_2003,Sheth_2003,Shandarin_2004,Bas2006,Weygaert_2011,Sousbie_2011,Park_2013}
%
confirming the picture
of a web-like network, but also to
study the intrinsic properties of the cosmic network, namely
connectivity and complexity
(cf. \citealt{Edelsbrunner_2002,vandeWeygaert2008,Arag_n_Calvo_2010,Aragon-Calv_2010b,Edelsbrunner_2010,Cautun_Weygaert_2014,Hong_Barabasi_2016,Pranav_2017,Libeskind_2017,codis_2018,Feldbrugge_2019,Kono_2020,Biagetti_2021,Wilding_2021}).  For a review see also \cite{Wasserman_2018} and
references therein.

Among the large body of the proposed computational methods
for investigating the intrinsic properties of the cosmic web,
it is worth noting that of
\citet{BARAB-PAPER} who have
used graph theory and simple models in order to produce
a graph-like image of the cosmic network.
In particular, these authors using  data provided by a subhalo catalog constructed  from the Illustris cosmological simulation \citep{Vogelsberger_2014}, defined  seven distinct network models for the construction of the 
cosmic web (M1-M7). M1 is the simplest of these models which links two nodes with an undirected link if the distance between them is less than a predefined length $l$. On the other hand M2 (and M3) is the directed version of M1 with the specification that each node is connected with a directed edge with its k nearest neighbors (M3 inverts the edge direction of M2). In M4 (and M5) the model inserts a directed link between two nodes if the distance between them is less than $a \cdot R_i^{1/2}$ (M5 inverts the edge direction of M4), where $R_i^{1/2}$ is the half-mass radius of the galaxy represented on node $i$ and $a$ is a predefined hyper-parameter constant. Model M6 (and M7) is a direct extension of M4 (and M5) which inserts a direct link if the sum of the square of the normalized distance and relative speed between two nodes (each one of them representing a galaxy) is less than $a^2$, where $a$ is again a hyper-parameter constant (M7 inverts the edge direction of M6).
It is assumed that subhalos in the simulation correspond to galaxies in the observational data representing the nodes of the cosmic web.
The authors   measure correlations between connected nodes as a function of various galaxy properties such as the star formation rate, finding the best model maximising the correlation (model M3). They
find that   model  M3 which relies only on spatial proximity offers the best correlations between the physical
characteristics of the connected galaxies.
To further validate M3, the authors compared the simulations with observational data taken from the Sloan Digital Sky Survey (SDSS) \citep{SDSS} showing  an excellent agreement between the in-degree distribution (fraction of nodes with a given in-degree), number of strongly connected components  and average clustering coefficient \citep{BARAB-BOOK}  between the observational and the simulated M3 networks.
However, \cite{SDSS} do not classify the objects detected  
as different cosmic structures such as voids, walls and clusters.
The latter is addressed in  \cite{Hong_2015} who used various simple network measures, such as the Degree Centrality (DC), Closeness Centrality (CL) and Betweenness Centrality (BC) \citep{BARAB-BOOK} to build a network of the cosmic web using data from the COSMOS catalog \citep{COSMOS_2013}. In particular, the authors utilized the value of each of those topological notions for the objects of the  COSMOS catalog to identify the kind of cosmic structure that object is. Using these network centrality notions, they identify eight distinct  structures: Void, wall, cluster (using the DC), main branch and Dangling leaf (using the BC), and Kernel, Backbone and Fracture (using the CL). Furthermore,   the authors investigate whether these topological classes identified are correlated to other properties of the galaxies, such as the star-formation-rate (SFR), the stellar mass, or its color (blue or red),  showing several interesting results.
%

In a paper that follows-up this work, \citet{Hong_Barabasi_2016} used  network analysis to  discriminate between topologically different distributions that have similar two-point correlations. In particular, they compared two galaxy distributions with similar two-point correlation statistics but different topologies, one derived from a cosmological simulation and the other from a L\'{e}vy walk \citep{Mandelbrot1975}. They find that the simulated galaxies and L\'{e}vy walks are statistically different in diameter, giant component, and transitivity measurements, which shows that L\'{e}vy walks fail to mimic the topologies of the distribution of the simulated galaxies. Using network analysis, they showed that one can discriminate between topologically different distributions that have similar two-point correlation statistics.

In an alternative approach,  \citet{SIMUL}
explored  deep generative models and in particular  Generative Adversarial Networks (GANs) \citep{Kingma_2014,Goodfellow_2014} to synthesize samples of the cosmic web.
Deep generative models are able to learn complex distributions
from a given set of data, and then generate new, statistically consistent data samples. Generative Adversarial Networks   create such a model by adopting an adversarial game setting between two   players, a generator and a discriminator. Deep generative models have also been used to generate astronomical images of galaxies \citep{Prabhat_2015,Schawinski_2017}.

A well-known and extensively used method in the field of the study of the cosmic web and its evolution is the  Delaunay Tessellation Field Estimator (DTFE) method \citep{Schaap2000,van_de_Weygaert_2008}. It is a well established mathematical method to reconstruct a volume-covering using continuous density and intensity fields from a discrete set of points in space. This method utilizes the position of the points along with Delaunay triangulations in order to create density fields which are needed in the study of galaxies and their formation.

In this paper we describe work done in the same spirit as the work described above, namely we
apply various computational methods on  data included in the database of the  IllustrisTNG  cosmological simulation \citep{Springel_2018, Pillepich_2018, Nelson_2018, Naiman_2018, Marinacci_2018} 
in order to extract properties of the cosmic web.
The aim of this paper is
not to argue which method
provides the best way to identify the cosmic web. Our goal is 
to allow the reader to understand the
differences between the different cosmic-web finders so that it will
be easier to compare different studies of the cosmic network in the
literature. Moreover, we suggest the use of computationally efficient spatial clustering algorithms in order to significantly speed up identification of cosmic web structures. We also hope that the paper will stimulate more detailed
follow–up studies to work towards a more unified view of this
topic.
\normalem
In particular, in this  work we have explored spatial clustering algorithms \citep{miller2009geographic} for the detection and analysis of the cosmic web.
 Spatial clustering  algorithms take as input a given set of points located in a space and find a  partitioning of the set 
into groups of high density, closely placed, groups of points, which are called, in the Data Mining domain \emph{clusters}.
 Clusters detected by a clustering algorithm applied on cosmological data correspond to clusters or superclusters depending on 
their size. To avoid confusion between the two possible meanings of the use of the word cluster, we use the term communities for the clusters obtained as a result of a clustering algorithm.
We have examined three spatial computational methods for detecting and characterizing various parts of 
the cosmic web, as  voids, walls, clusters and superclusters. The methods are able to reveal the structure of the cosmic web in various 
resolutions and to use a variety of physical properties, such as internal density, size, distance and  mass. 

Their main advantage is that they can be accomplished in much faster time, making them suitable for use for larger databases of galaxies. 
This nice property allows one of them to be combined with the DTFE method to get results of similar quality but in one order of magnitude less computation time.  


This paper is organised as follows: in Section 2 we briefly present the simulation from which the test region was extracted.
In Section 3 we describe the spatial  methods utilized  and in Section 4 we describe and discuss the results obtained by the application of the methods on the simulated data. Finally, Section 5 provides our discussion and conclusions of this work and ideas for future work.

\section{Simulated Data Set}


The publicly available 
IllustrisTNG simulated data \citep{Springel_2018, Pillepich_2018, Nelson_2018, Naiman_2018, Marinacci_2018} 
is an ambitious suite of new hydrodynamical simulations of
galaxy formation in large cosmological volumes. The simulation database contains three simulation volumes, TNG50, TNG100 and TNG300. Each one of these volumes is available in three different levels of resolution (1, 2 and 3, where 1 represents the highest resolution level). All of the simulations assume a $\Lambda$CDM cosmological model with parameters tuned according to the Planck constraints \citep{Planck_Collaboration}: $\Omega_m = 0.3089, \Omega_b = 0.0486, \Omega_{\Lambda} = 0.6911, \sigma_8 = 0.8159, H_0 = 67.74$km s$^{-1}$Mpc$^{-1}$ and $n_s = 0.9667$. In this work we analyze the structure formations that are present in the TNG100-1-Dark periodic volume. The TNG100-1-Dark volume has a side length of 110.7 Mpc and a dark matter resolution of $8.9 \times 10^6 M_{\odot}$. The simulation was performed with the moving-mesh code AREPO \citep{Springel_2010} and uses the magneto-hydrodynamics equations for its evolution. IllustrisTNG is an updated version of the Illustris model \citep{ILLUSTRIS-2}. This new model uses self-gravity, magnetohydrodynamics, radiative cooling, star-formation, and an updated set of sub-grid physics models for stellar evolution, black hole and stellar growth, a new mode of AGN feedback scheme and numerical improvements for the convergence properties \citep{Weinberger_2017, Pillepich_2018}.\\
Haloes were identified via a standard friends-of-friends
(\textit{FoF}) algorithm with a linking length of b = 0.2 and substructures were identified using \textit{SUBFIND}. In this article we use dark matter subhalos of all masses given by the simulation data. We use data offered in multiple redshifts in order to study the chronological evolution and solidify the results of our proposed methods.

\section{Methodology}
In this section  we present the methods that we utilize, adjust or modify  for detecting and clustering extragalactic structures.
\subsection{Gravity Lattice} 

We have implemented a 3D lattice on the cosmic cube where each vertex of the lattice represents a 1kg gravitational test load. After that we have inserted all galaxies in their corresponding positions according to the data set \citep{ILLUSTRIS-2} and calculated the gravitational forces they exerted to the test loads inside their field of influence. Finally, we move the test loads with respect to the forces that have been exerted over them. The algorithm's time complexity is $O( c \!\cdot \!n )$, where $c$ is the maximum number of test loads within the range of influence of each galaxy, which is bounded by a constant,
and $n$ is the number of data points. The algorithm was further improved using a smart filtering on the test loads based on the value defining how the test loads have been moved in the previous procedure (we have moved them according to the total forces exerted on them from the galaxies) and the number of galaxies that have affected them. This filtering allowed us to examine galactic structures of predefined size, which is  useful information for the understanding of the various structures of the cosmic web. The aim behind this model is to extract a gravitational footprint of the galaxies and then filter that footprint accordingly\footnotemark[1]\footnotetext[1]{By `footprint', we mean that each galaxy leaves a gravitational mark on the fabric of the cosmic web (based on its mass) which we utilize in order to detect cosmological structures.}.
The intuition behind the filtering is based on the astrophysical properties: Galaxies in void regions are few or none and as a result the test loads that lay on these regions will have a small number of galaxies that will affect them. On the other hand, test loads in galactic clusters or superclusters will be affected by more galaxies. Filtering on the number of galaxies that affect each test load results in different levels of resolution of the filter. These different values filter out either smaller or larger galactic structures allowing us to define \emph{voids} (test loads are affected by a small number of galaxies), \emph{walls} (test loads are affected by a moderate number of galaxies), \emph{clusters} (test loads are affected by a large number of galaxies) and \emph{superclusters} (test loads are affected by a very large number of galaxies),  depending on the number of the galaxies these structures affect.

\subsection{ABACUS}
We  modified the spatial clustering algorithm ABACUS  \citep{ABACUS} for detecting and categorizing various galactic structures of the cosmic web. The ABACUS algorithm is a spatial clustering algorithm that detects  arbitrarily shaped clusters (i.e. communities) in data placed in a given space.   ABACUS is based on the idea of identifying the intrinsic structure, i.e. the \emph{backbone}  for each cluster by merging and moving nearby points.  Specifically, the algorithm selects suitably a small fraction  of the initial points  (called the \emph{representative} points) to form  a  \emph{backbone},   representing the main structures of the original points. Through merging the original points according to a merging radius chosen and  moving the representatives points of the backbone accordingly, the algorithm manages to detect communities of the backbone points or clusters as called in \citet{ABACUS}  of  high  density, closely located points. We note
though that the points detected by the algorithm are not the same as the original ones. Reversing the procedure could retrieve the original points while keeping their association to the backbone points.

\begin{figure*}
    \centering
    \includegraphics[width=0.9\textwidth]{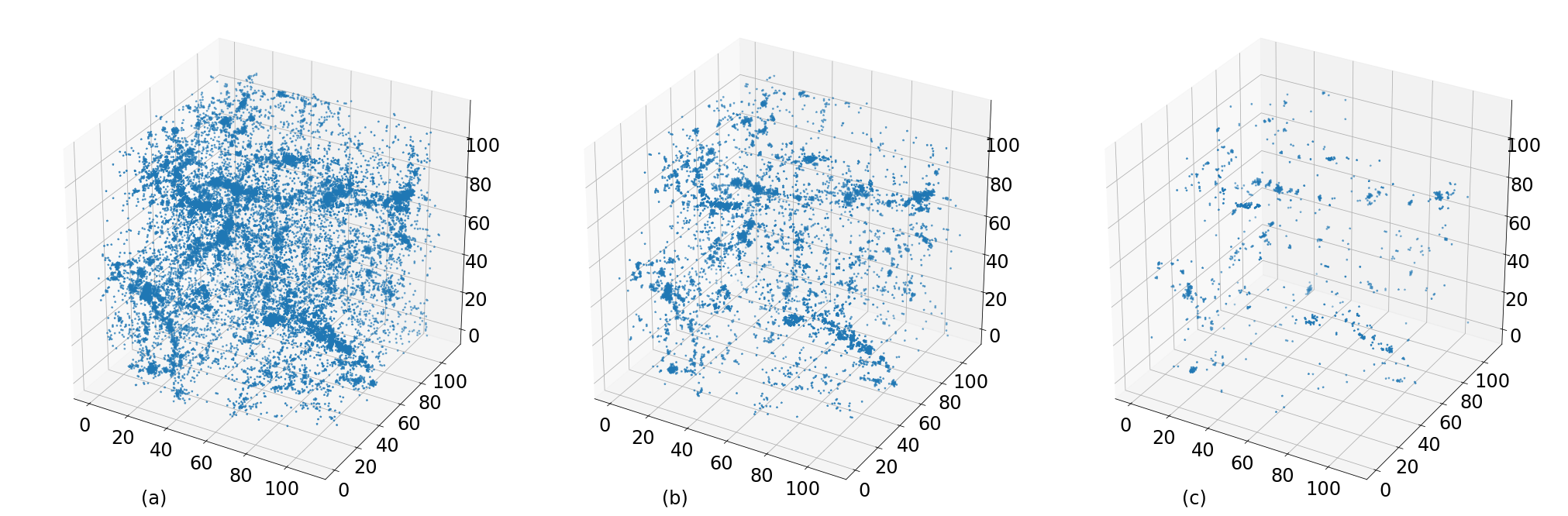}
    \caption{Gravity Lattice Results. Each cube has an edge of ca. 110 Mpc. We applied Gravity Lattice with a regular 3D lattice where the distance between the test loads is 0.5 Mpc/h. Then we applied a filtering based on the number of galaxies affecting each test load. The filter rules out test loads affected by less than the threshold number of galaxies. (a): Threshold = 10, (b): Threshold = 20, (c): Threshold = 50. We observe that  the increase of  value of the  filter threshold enables the filtering of the detected cosmic web to only  the larger structures within the cosmic cube. The units of the axes in this and all following figures are Mpc.}
\label{fg:Grav_lattice_results}
\end{figure*}

Here, we utilize only the first part of the algorithm, i.e. the part of the detection of the backbone of the set of points considered, in order to get the main structure of the cosmic web. Then,  a  suitable filtering  on the detected backbone allows us to reduce the size of the input to DTFE, while maintaining a substantial fraction of the original information that lies in the galaxies' positions. As a subsequent step, using the above-mentioned backbone could directly
enable the detection of  galactic structures of various sizes (in terms of the number of galaxies included in each such structure).
Additionally, the algorithm has been enriched so that the backbone  structure   detected    (which in fact represents more that one of the initial points),
stores the total mass instead of the number of the initial points that it represents. This modification enables then an alternative filtering,  according to the mass of the initial points, which in turn enables the extraction of galactic structures of different masses.

We also note that   ABACUS utilizes  $k$-Nearest Neighbours-graph  ($k$-NN) \cite{Preparata-k-NN}  (for considering only the $k$ nearest neighbours of each point)   and by tuning $k$ accordingly we are able to focus on the local neighborhood or extend to further neighbors of every galaxy. Increasing the value of $k$ results in extended neighborhoods; thus, giving different results for the formation of the detected cosmic web. This is  because the glob-move procedure happens in a different manner, thus the number of resulting points in the backbone are differently placed and more points are obtained. On the other hand, holding a relatively small value in $k$ results in paying attention to the close neighborhood of each galaxy. We  believe that this is more suitable for  the study of the cosmic web since the focus is on the  relatively closest neighbors of each galaxy. Finally note that an  increase of  the value of $k$ results to an analogous increase of  the computation time  of   the ABACUS   algorithm.

Regarding the time efficiency of the modified algorithm, it can be easily seen that the algorithm is quite  fast, i.e  of   time  complexity $O(n \log n)$, where $n$ is the number of points. The slowest part of the algorithm is the computation of the $k$-NN-graph needed, which requires  O($n \log n$) time.

We note that  the ABACUS algorithm can be applied as a pre-processing step for reducing the number of sample points which can then be used as  input for the DTFE algorithm reducing the computational time while maintaining the quality of the results. More details are explained in Section 4.
Finally, we  observe   that the ability of the algorithm to detect the backbone of the system under investigation could be useful for the analysis of the cosmic web since it reveals a minimum structure of it which could actually have appeared at some point during the evolution of the universe.

\subsection{HDBSCAN}
We utilize the well known spatial clustering algorithm HDBSCAN  \citep{HDBSCAN} with suitable fine tuning of the values of its parameters to detect the cosmic web,  highly dense structures within it (communities) and utilizing the over-density (shown in Equation \ref{eq:1}) we have managed to categorize various parts of it as voids, walls, clusters and superclusters.
\begin{equation}\label{eq:1}
    \hspace{2.7cm}
    \delta(x, t) = \frac{\rho(x, t) - \rho_u(t)}{\rho_u(t)}
\end{equation}
where $\rho$ are the densities from DTFE and $\rho_u$ is the mean density at that particular epoch.
More specifically, the HDBSCAN algorithm is applied on an input space of data points, using   three parameters (taken also as input) to   partition  the data set   into a set of high density regions of closely placed points,  called in \cite{HDBSCAN} as clusters and  here called \emph{communities}. The three parameters used by the algorithm are (\emph{i}) the minimum number of points per community, (\emph{ii}) the maximum intra-community distance (maximum distance of points in the same community) beyond which a community tears apart into two different communities and (\emph{iii})   the percentage of detected noise by the algorithm.

We apply the HDBSCAN algorithm on the simulations under investigation, choosing appropriate values of these three parameters using  domain specific knowledge. In order to be able to  categorize detected communities  as  voids, filaments,  clusters or superclusters we make use of the over-density factor within each of the detected communities.
The careful choice  of those three parameters enables not only the detection of the cosmic web but also the categorization and clustering (i.e. partitioning) of various parts of it into various kinds of galactic structures, i.e., voids, walls, clusters and superclusters.  In particular, with fine tuning of the parameters and  exploiting the ability of the  HDBSCAN algorithm to   detect which points of the input are actually \emph{noise}, we  detect the parts of the detected cosmic web which correspond to voids.  Additionally, the suitable choice of the value of the parameter for the minimum community size to be detected,  enables the  detection of   galactic structures of various sizes. Finally,  varying the value of the   parameter for the distance   beyond which  to separate communities, allows the  detection of  communities of various scales, that is, communities partitioning the  whole data set as well as communities within the detected communities (sub-communities). We note that not all communities detected correspond to clusters or superclusters; they constitute either dense regions of the cosmic web that could be walls, filaments or clusters or sparse points of the original dataset which represent void regions. However, this is finally accomplished by  applying suitable values of the input parameters of the algorithm.

\begin{figure*}
    \centering
    \includegraphics[width=0.9\textwidth]{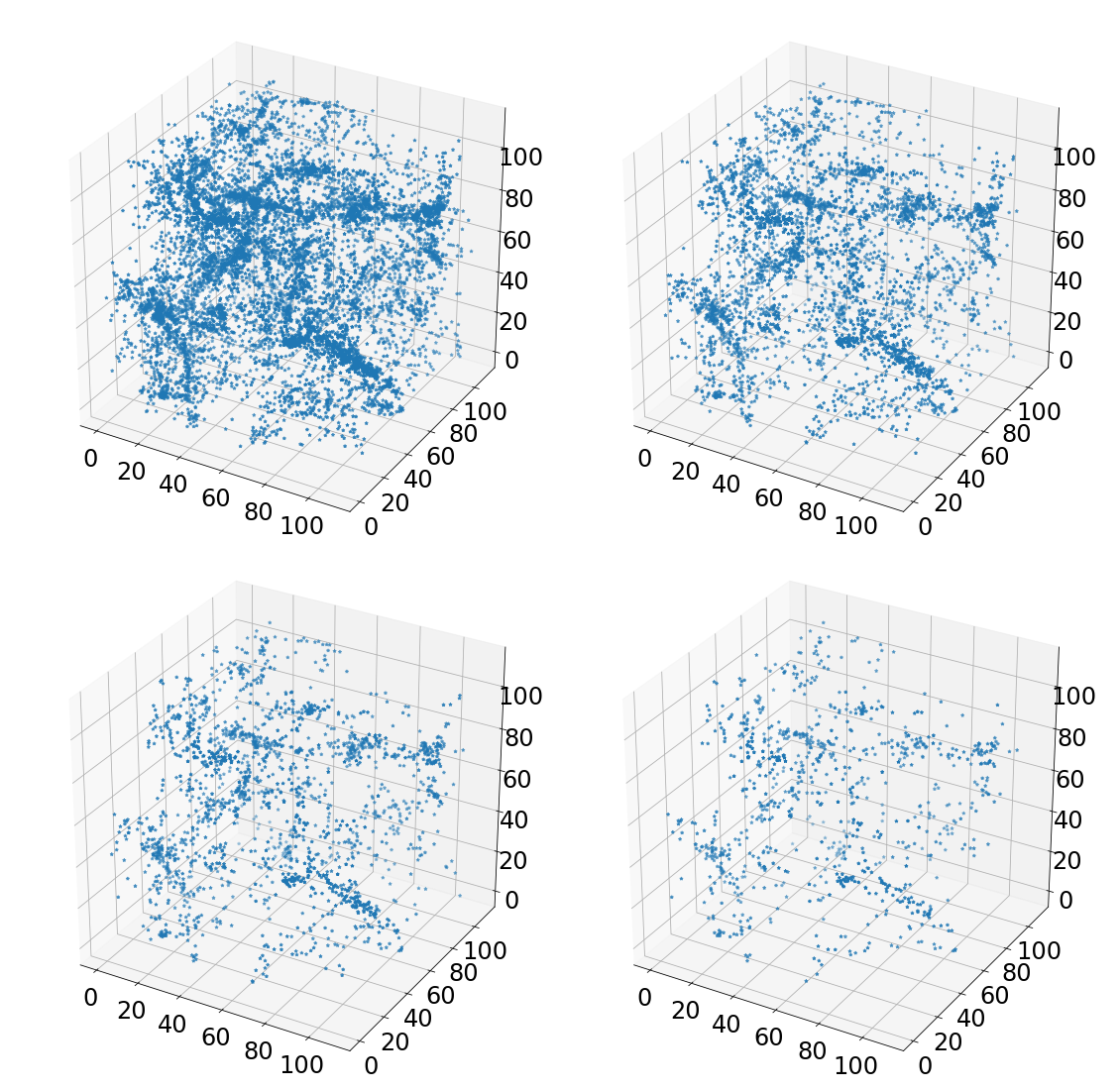}
    \caption{The cosmic web obtained using the ABACUS method and after filtering over the number of galaxies  each final point represents  using a   $3-NN$ network. Each cube has an edge of ca. 110 Mpc. The filtering method we applied rules out points that represent fewer galaxies than the threshold value. We show  the  detected cosmic web for different values of the threshold. On the upper row we show the results for thresholds $3$ ,$4$, (from left to right) and on the bottom row we have the results for thresholds $5$ and  $6$.}
    \label{fg:ABACUS_results_points}
\end{figure*}

\begin{figure*}
    \centering
    \includegraphics[width=0.9\textwidth]{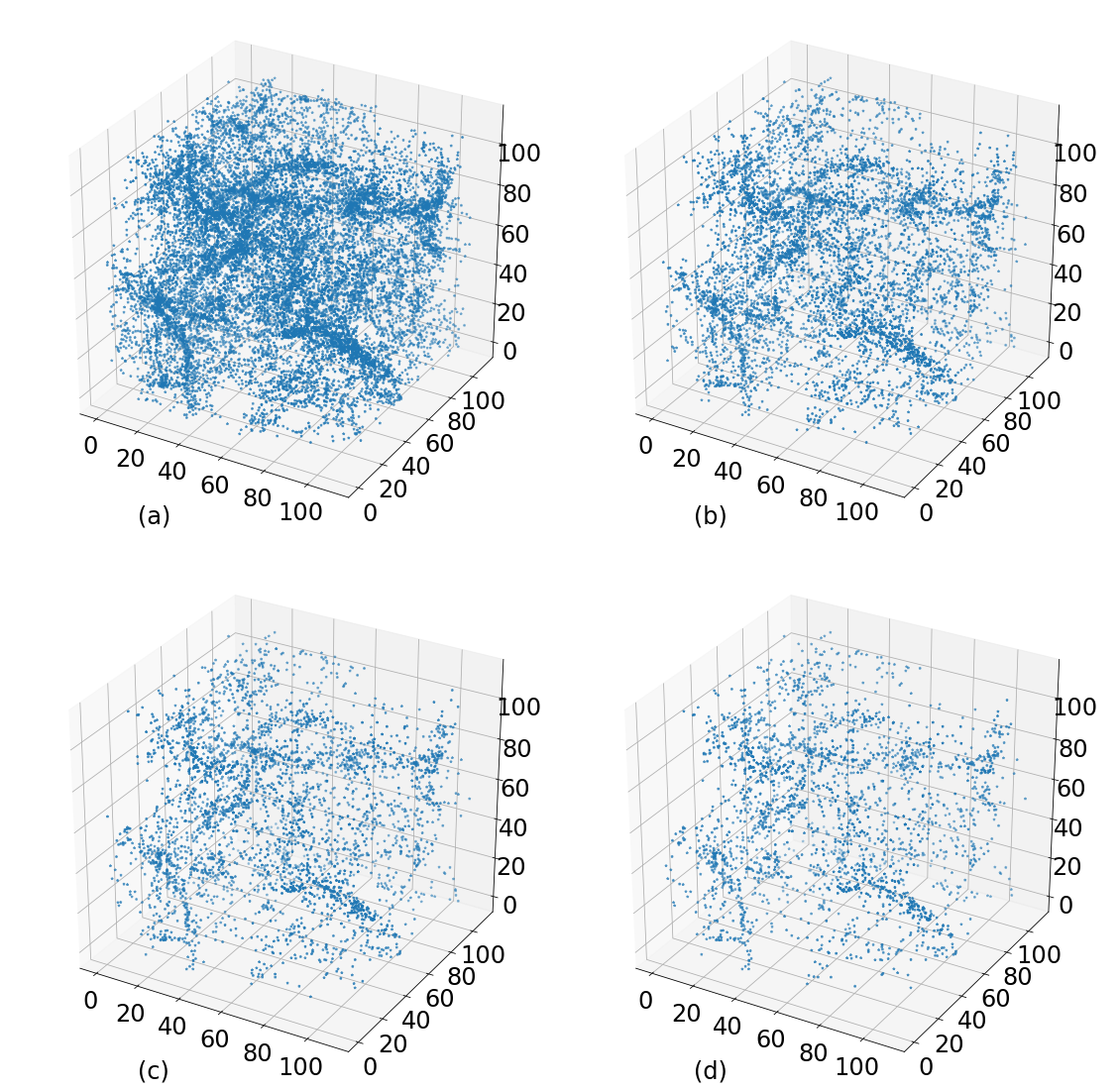}
    \caption{The cosmic web obtained using the ABACUS method and after filtering over the total mass of galaxies each final point represents  using a   $3-NN$ network. Each cube has an edge of ca. 110 Mpc. The filtering we have applied rules out points that represent less total mass than the threshold value. We show the cosmic web detected with different values of the threshold. All masses and filter values are measured in units of $10^{10} M_\odot$. On the upper row we have the results for thresholds $10$, $30$ (subfigures (a) and (b)) and on the bottom row we have the results for thresholds $50$, $70$ (subfigures (c) and (d)).}\label{fg:ABACUS_results_masses}
\end{figure*}

\subsection{DTFE Method}
 In this paper we also utilize the DTFE method \citep{Schaap2000,van_de_Weygaert_2008},   a mathematical   method implemented  in publicly  available C++ code \citep{cautun2019dtfe},  for   reconstructing  a volume-covering and continuous density and intensity fields from a discrete set of samples/measurements, using the maximum of information contained in the point
distribution.  This interpolation  method offers  several   advantages:  (\emph{i}) it automatically adapts to strong variations in density and geometry.  It is therefore very well suited for studies of the large scale galaxy distribution.  (\emph{ii})
It preserves the local geometry of the point distribution making it suitable for detecting  sharp and anisotropic features like the different components of the  large scale structure (i.e. clusters, filaments, walls and voids) and (\emph{iii}) the  interpolated fields are volume weighted.

\section{Results}

In this section we present the results obtained from the application of each of the three methods investigated: The Gravity Lattice, the ABACUS and the HDBSCAN method. In the last part of the section, we show the results of HDBSCAN equipped with the proposed usage of over-density to categorize the cosmic structures and
finally we combine DTFE with the ABACUS method and demonstrate that the combined method can achieve similar accuracy with significantly less time than the original DTFE method.

\subsection{Gravity Lattice}

Using the Gravity Lattice formation method, we have conducted experiments over the 3D cosmic cube and filtered the resulting test loads on the number of the galaxies that have affected them. While inserting each galaxy to the Gravity Lattice we keep track for each load the number of galaxies that have affected it. On that number, we perform a filtering on the three dimensions and we keep the points affected by at least a number of galaxies as specified by a threshold. The cosmic web detected in this way is shown in Figure \ref{fg:Grav_lattice_results}.

We note that  by construction, test loads are expected  to `collect' a number of nearby galaxies  (which have affected each test load), resulting to a more compact representation of the  set of  points (galaxies) compared to the original data set. Hence, this method can be viewed as  a down-sampling method of the whole galaxy-position set to a smaller more compact one. Furthermore, note that according to this method,  we do not obtain the original  positions of the points (galaxies) but the  positions of the test loads.

Furthermore,  filtering on the number of galaxies  affecting the test loads enables  detection of smaller or larger scale structures, accordingly. We have also performed filtering based on the mass of the galaxies that affect each test load resulting in similar yet less effective results as the results from filtering on the number of galaxies.  We note   that the method can be applied as a pre- or post-processing step of the DTFE method, to   capture more large or small scale structures of the cosmic web, through suitable  tuning of the distances between test loads.
Finally, we note that   Gravity Lattice can also be used in the 2D projection of a slice of the cosmic cube and    perform similar  analysis. The presented results are preliminary and we would like to leave further investigation of the method as future work.

\begin{figure*}
    \centering
    \includegraphics[width=0.9\textwidth]{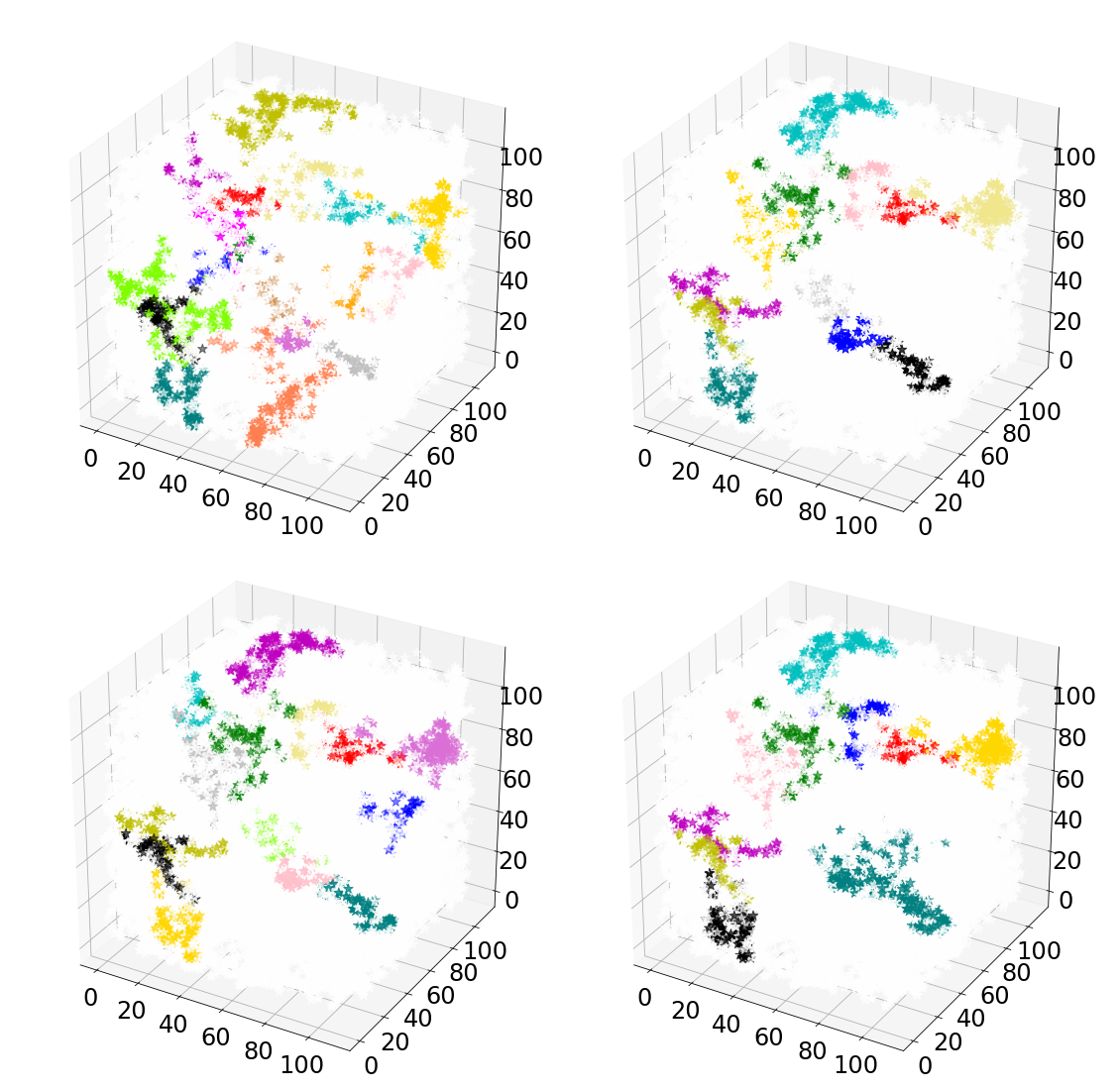}
    \caption{The cosmic web and its decomposition into  communities   using the HDBSCAN algorithm. Each cube has an edge of ca. 110 Mpc. Each result is  associated with a triplet  [MinClusterSize, MinSamples, Epsilon], corresponding to the parameters (\emph{i})- (\emph{iii}) specified in Section \ref{HDBSCAN_results_section}.
    In the Upper row we plot the results for: [700, 30, 1], [1200, 100, 1], and in the Lower row: [900, 100, 1],
[1500, 100, 1]. We observe that in all  setups the algorithm manages to find communities of different formations and sizes.  Additionally, we observe that communities detected with this method can also be further explored using HDBSCAN in an iterative manner with different parameter tuning resulting in smaller communities existing within the bigger ones.}
\label{fg:HDBSCAN_results}
\end{figure*}

\subsection{ABACUS}
In this subsection we present the results obtained using the modified  ABACUS method with appropriate  filtering. The results obtained seem to eliminate  almost all void galaxies which can be used for determining void regions. We have used this algorithm and filtered out some lightweight point of the resulting backbone. The rational for this is that the lighter the point, the closer to other points located nearby and as a result it represents either only itself or a few galaxies. This is probably because it lies on empty regions with no or very few galaxies (which are the void structures). On the other hand, points (galaxies) that lie inside denser regions of clusters or superclusters end up to represent many points and have more mass. Filtering the results of ABACUS with different thresholds (of both the original and the modified algorithm where  weight is used to represent total mass) allowed us to extract different sizes of galactic structures. Although the resulting points move from their original positions, as a result of the application of the algorithm, the backbone structure is maintained  and the final positions of the points  represent the backbone of the galactic structure which in fact is what we need in order to categorize a region as void or cluster/wall. Void regions are the regions that end up to have no points after the filtering while the cluster/wall regions are the regions that have points after filtering as well as the regions around them. We may also consider these regions as walls, because the algorithm has moved the initial points. These classifications are rough and will not be used in the subsequent analysis. We rather note them as initiatives to explain the ABACUS power, the intuition behind our method and why we have used ABACUS as a DTFE preprocessing method.

The cosmic web formations  obtained using  the modified ABACUS method
through various values of filtering, are shown in Figures \ref{fg:ABACUS_results_points}   and  \ref{fg:ABACUS_results_masses}. We have conducted experiments where we have varied the value of $k$ in the $K$-NN network in the range $3$ to $10$;  we  present the results of the (modified) ABACUS algorithm using a $3$-NN network. We have chosen these results as they appear more clear and interesting for understanding the cosmic web.

In Figure \ref{fg:ABACUS_results_points} we  show the  cosmic web detected  after applying the ABACUS method. We first observe that the resulting points of ABACUS are fewer than the original number of points which is equal  to the number of galaxies. This is because ABACUS merges points and this results in a smaller number of points than  the original points. Sometimes the resulting points may even be fewer than half of the original points. Figure \ref{fg:ABACUS_results_points} demonstrates  how different levels of filtering  result in the detection of different kinds of galactic structures  of the   cosmic web detected by the algorithm. Filtering on higher values, results in  keeping only points of denser regions  and at the same time filtering out sparse regions of the cosmic web. The algorithm generally reduces the number of   points that we keep to  only points that represent more galaxies than the threshold value. As we increase the threshold the number of points diminish, revealing  the regions of high density of the cosmic web.

In Figure \ref{fg:ABACUS_results_masses} we show the results of modified ABACUS where each of the resulting points have the total mass of the galaxies it represents. In this setup, we have performed a filtering, so that to keep only the  points that represent total mass of value at least   the value of the  threshold used in the filtering. The rational for this is that points located in dense regions will represent more galaxies and thus will be more heavy than points located in empty regions. With this filtering we aim to filter out points in voids and keep only points inside galactic structures in an attempt to distinguish voids from clusters.

\begin{figure*}
 \centering
    \includegraphics[width=0.9\textwidth, scale=0.95]{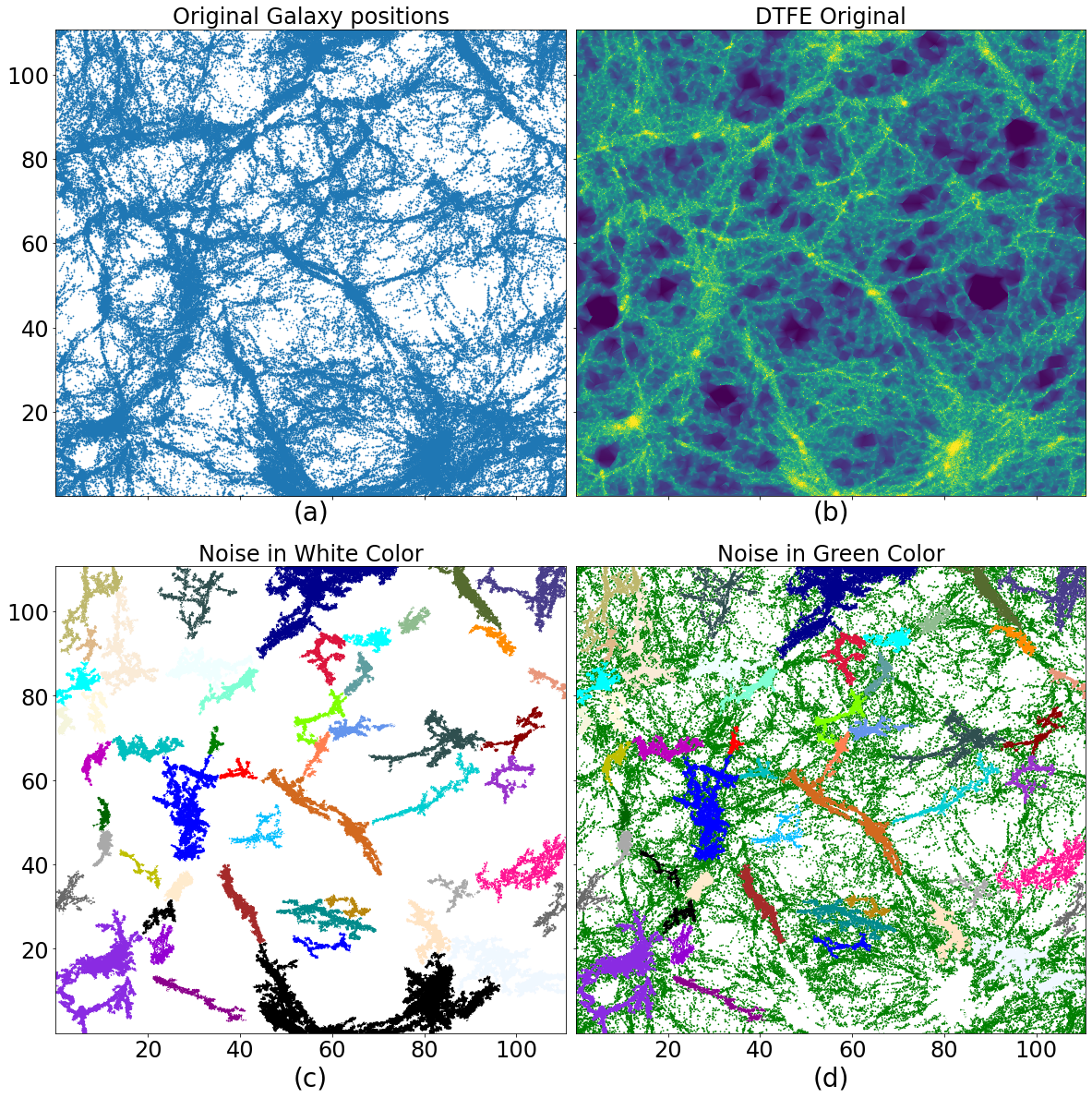}
    \caption{Comparison of the extended HDBSCAN and DTFE results. HDBSCAN is parameterized with MinClusterSize = 500, MinSamples = 10 and Epsilon = 0.35 Mpc/h. Upper Left (a): Original positions of ca. 232k galaxies. Upper Right (b): DTFE density results over galaxies' positions.
    Bottom Left (c): Results from HDBSCAN (noise with white and parameterized as defined above). Bottom Right (d): Results from HDBSCAN (noise with green and same parameters as above).}
    \label{fg:DTFE_plus_HDBSCAN}
\end{figure*}

\begin{figure*}
    \centering
    \includegraphics[width=0.9\textwidth, scale=0.95]{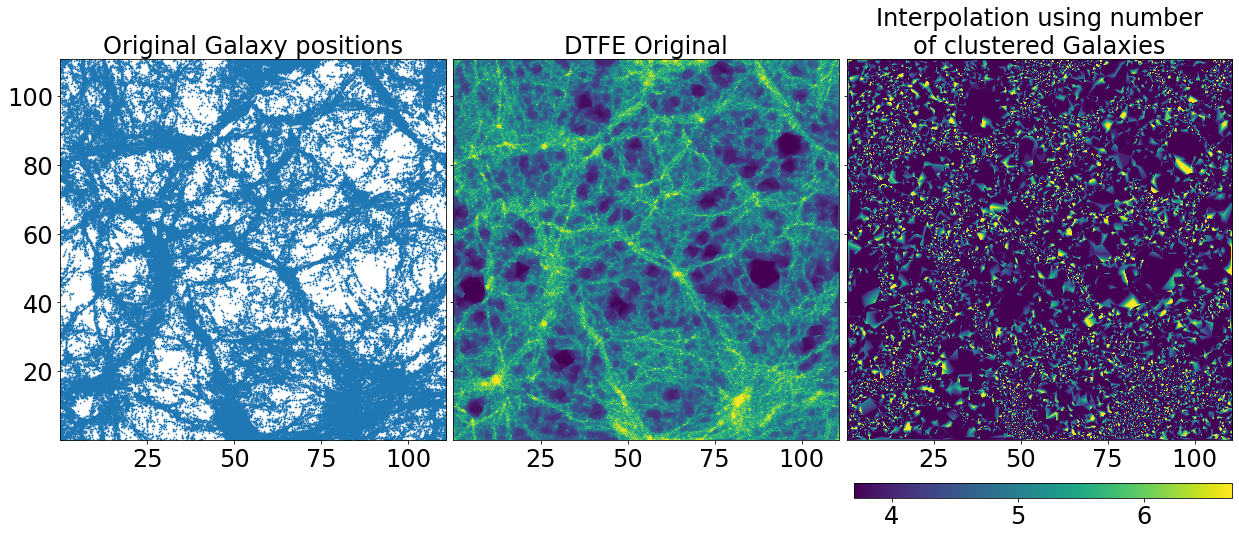}
    \caption{Left: Original positions of ca. 232k galaxies. Middle: DTFE result for the same galaxies Right: Results from interpolation over the ABACUS weight. By interpolation we mean that we first run the ABACUS algorithm and each of the resulting points has a weight depicting the total number of galaxies this point represents. In a DTFE-like approach we interpolate that weight in order to produce the final image. On that image, points with brighter colors show that these points represent more galaxies. Points in darker colors represent fewer or no galaxies. This attempt is inspired by DTFE and we try to find the density through the number of galaxies each of the final points represents. }\label{fg:ABACUS_interpolation}
\end{figure*}

\subsection{HDBSCAN}\label{HDBSCAN_results_section}

Using the  modified version of the HDBSCAN method we manage to detect and  categorize various parts of the cosmic web and also to  detect 
highly dense regions (i.e. communities) within it.
The results are shown in Figure \ref{fg:HDBSCAN_results}. We have tested various values for each one of the  three parameters of the HDBSCAN algorithm aiming to find the best and most appropriate combination  for the problem studied. Recall that 
  HDBSCAN takes  three input parameters: (\emph{i})  the minimum number of points per community, (\emph{ii}) the percentage of points that will be considered  as noise and (\emph{iii})   the  distance beyond which the algorithm splits a community into two smaller communities. Figure   \ref{fg:HDBSCAN_results} shows the detected cosmic web and  communities within it on the four most promising setups for the above mentioned parameters.

 We observe that in all of the results presented, the noise (as detected by HDBSCAN and noted with white color)  covers a large area of the cosmic web. By construction, the regions where the HDBSCAN algorithm detects noise correspond to voids in the cosmic web. Additionally, the various communities detected by the algorithm correspond to various kinds of cosmic structures, as we explain next.

 HDBSCAN is able to detect 3D structures of different sizes and shapes in the cube and capture relations between galaxies based on their distance. As we can observe in all the results shown in \ref{fg:HDBSCAN_results}, there exist both small  and large  communities. We are able to capture the density of the cosmic web and illustrate complex formations. In all four schemes the  cluster minimum distance is equal to 1Mpc. The algorithm therefore splits  a community into two smaller  communities if its diameter is larger than 1Mpc. The  value has been chosen to  be  equal to the    theoretical value of the diameter of a galactic cluster/supercluster. Moreover, we vary the minimum number of galaxies per cluster, obtaining in this way, communities of various minimum size.  Changing the size of each community results in a different size of the resolution per community. Smaller communities may be located inside larger ones. We can also observe that, although the algorithm runs using different parameters the resulting communities appear to have some similarities in size and location. This suggests that the algorithm can detect communities of different sizes which might be sub-communities of larger communities in a different setup. Each scheme on Figure \ref{fg:HDBSCAN_results} is associated with   a triplet of the parameters of HDBSCAN as [MinClusterSize, MinSamples, Epsilon], where   MinClusterSize specifies, the value of  the minimum number of point per community, MinSamples is the percentage of points that will be considered  as noise and Epsilon  is    the  distance beyond which the algorithm splits a community into two smaller communities.

\begin{figure*}
    \centering
    \includegraphics[width=0.9\textwidth]{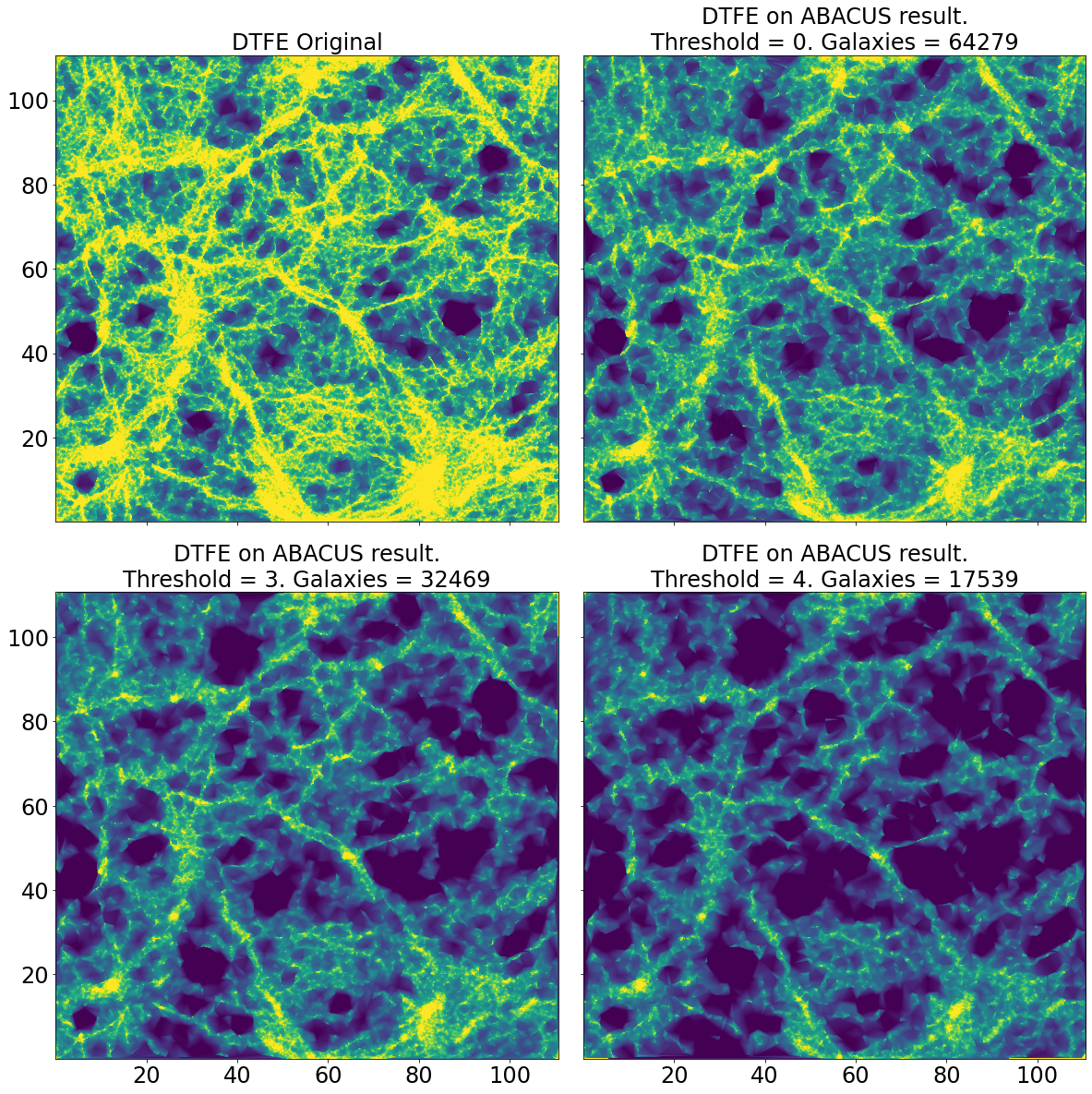}
    \caption{Here we show the results of applying DTFE on the ABACUS resulting points after we filter them based on the number of galaxies they represent. Each point has a weight showing the number of galaxies globbed on this point. Upper Left: Original positions of ca. 232k galaxies. Upper Right: DTFE over ABACUS results.
    Bottom Left: DTFE over ABACUS results after filtering out points with weight less than 3. Bottom Right: DTFE over ABACUS results after filtering out points with weight less than $4$. In this approach we try to use ABACUS as a down-sampling pre-processing method before calculating DTFE because DTFE is time consuming. As we can see in the above images, applying ABACUS filtering over ABACUS results and then calculating DTFE results in similar density fields of the cosmic web yet being time efficient as we calculate DTFE over a fraction of the original galaxies. A characteristic of this method is that the upper left and the lower right image look very similar while the latter uses only 10\% of the galaxies while the first uses the whole galaxy set.}\label{fg:ABACUS_DTFE}
\end{figure*}

\subsection{Methods' Results}

\subsubsection{HDBSCAN equipped with over-density}
Here,  we   examine the results of HDBSCAN and observe possible similarities with the classical DTFE approach both in time complexity and quality of results.
We first note that the classical DTFE uses Delaunay tessellations in order to construct its final representation, resulting in required  O($n^2$) time, where $n$ is the number of points. On the other hand, HDBSCAN outperforms  DTFE in terms of time required,  since it takes   $O(n\log n)$ time. To enlighten the reader about the magnitude of the speedup, for a task that takes one hour to be completed using DTFE, HDBSCAN manages to finish the computation within $\sim$ four minutes for the same task.

The classical DTFE method shows  the   structure of the cosmic web by highlighting the high-density structures and decaying this color as we move to less dense structures. On the other hand, HDBSCAN, using the same criterion (densities), detects, but also groups together (into different  communities of different color), various detected structures based on both positioning and density criteria. The results of the application of both methods is shown in Figure
\ref{fg:DTFE_plus_HDBSCAN}.
  Figure \ref{fg:DTFE_plus_HDBSCAN} shows galaxies on a slice of 5 Mpc/h around the 52 Mpc/h where according to the simulation there exist 232k galaxies. On the top row we may see their initial positions and the resulting DTFE while on the bottom row we see the results using HDBSCAN.

On the other hand,   DTFE  can   detect  various density levels of the detected structures    (shown with various brightness levels), while HDBSCAN does not (at least directly). However,  through the detection of (high density) communities obtained by    HDBSCAN, the   brighter areas of DTFE are actually detected, as also shown in Figure \ref{fg:DTFE_plus_HDBSCAN}.
At the same time, lower density structures, shown in DTFE with darker color, are also detected and are all colored with a distinct color in HDBSCAN (green or white color in   Figure \ref{fg:DTFE_plus_HDBSCAN}(c) and (d)),   distinguishing in this way, void galactic structures from higher density galactic structures.

As a result, HDBSCAN  manages to  detect the cosmic web,   decompose it into high density  communities (each colored    differently)  but also to  distinguish these structures from the rest of the cosmic web which is of low density (and color it with another single color), and is treated as noise.
The bottom row of Figure \ref{fg:DTFE_plus_HDBSCAN}  shows the results.  On the left, the noise is shown with white color while on the right with green. In this way, one can see both the detected noisy (sparse) regions and also the ability of the  algorithm to distinguish them from the high density ones, which are further decomposed into high density communities.

\begin{figure*}
    \centering
    \includegraphics[width=0.7\textwidth]{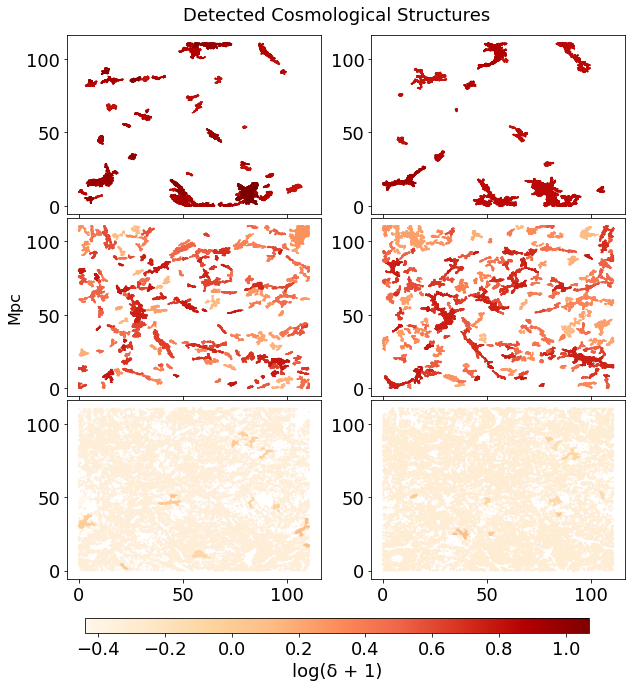}
    \caption{Detected structures from the application of HDBSCAN equipped with the over-density knowledge in order to categorize each community. The left column is for Redshift $z = 0$ and the right for $z = 1$. The  top row   shows the highest over-density structures corresponding to   Clusters. The middle row shows the middle over-density structures, corresponding to Filaments and Tunnels. Finally, the bottom row shows the lowest density structures, corresponding to  Walls \& Voids structures.}\label{fg:detected_structures}
\end{figure*}

\begin{figure*}
    \centering
    \includegraphics[width=0.5\textwidth]{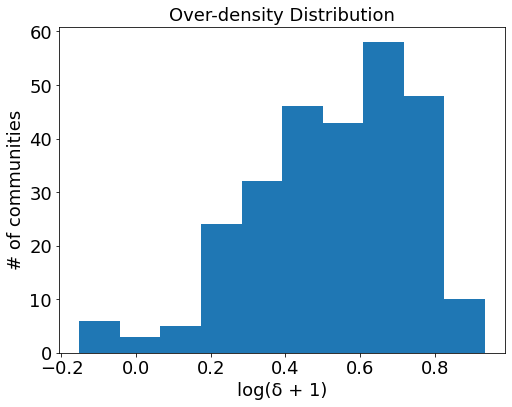}
    \caption{Over-density distribution. The x axis shows $\log(\delta + 1)$ and the y axis shows the number of the detected communities with such over-density value.}\label{fg:distribution}
\end{figure*}

To further categorize the detected communities we made use of the over-density notation. First we have performed a triangulation on the total set of data points (i.e. galaxies positions). For each of the detected communities we consider only the triangles having all of their vertices within the community and calculate their total surface area. Following the intuition of the over-density, we compute it per community using the number of community members and the surface of the respective triangles. Afterwards, following the notations of \cite{Wilding_2021}, 
we have created the distribution of $\log(\delta + 1)$, where $\delta$ is the over-density, as shown in Figure \ref{fg:distribution}. Based on that distribution, we have carefully selected thresholds in order to classify the galactic structures. We performed this procedure for two different redshifts, $z=0$ and $z=1$.  
Figure~\ref{fg:detected_structures}  provides a visual presentation of the  various cosmic structures detected,  based on three different density levels, corresponding to  the selected thresholds   obtained by the   over-density distribution (Figure \ref{fg:distribution}). 
Additionally, the first column of the  Figure shows the detected cosmic structures  for $z=0$ and the second column for  $z=1$.    For the representation of our results we follow a similar manner as the authors in \cite{Wilding_2021}.

The top row of Figure \ref{fg:detected_structures} shows the detected clusters (highest over-density) in two redshifts, the middle row the detected Filaments and Tunnels (medium over-density) and the last row the Walls and Voids (low over-density). One can observe an increase in both the low and moderate over-density structures (Walls and Voids, and Filaments and Tunnels) at the higher redshift. In contrast,  one can observe fewer but of larger size clusters (highest density)    at higher redshift.

In particular, the top row shows the structures at the highest threshold, at which level we observe the presence of high density structures. Following the gravitational evolution we observe that for smaller redshifts there exist more and larger high-density structures. On the contrary, for early ages (i.e. z = 1) we can see fewer and less dense structures, which is in accordance to the gravitational evolution of the cosmos.
The middle row shows the structures at the intermediate threshold, at which level we observe the presence  of filaments and Tunnels. These patterns  seem to increase both in density and in number moving from low to high redshift. This is due to the fact that some of them are going to evolve into high-density structures. Excluding that fraction of structures we can see that the rest of the Filaments \& Tunnels are fewer in earlier epochs and of less density.

The bottom row shows the structures at the lowest  threshold, corresponding to  Walls \& Voids. These patterns seem to decrease both in density and in number moving from high to low redshift. The aforementioned observation indicates, that in low redshifts there are more under-dense and empty regions, which is expected based on the gravitational evolution of the cosmos.

Finally, we observe that 
our method manages to bring together clustering methods (HDBSCAN) with existing methodology in the astrophysics domain (over-density). The results seem promising while maintaining similarities with the results of \cite{Wilding_2021}

\subsubsection{ABACUS results in DTFE-like manner}

\begin{figure*}
    \centering
    \includegraphics[width=0.9\textwidth]{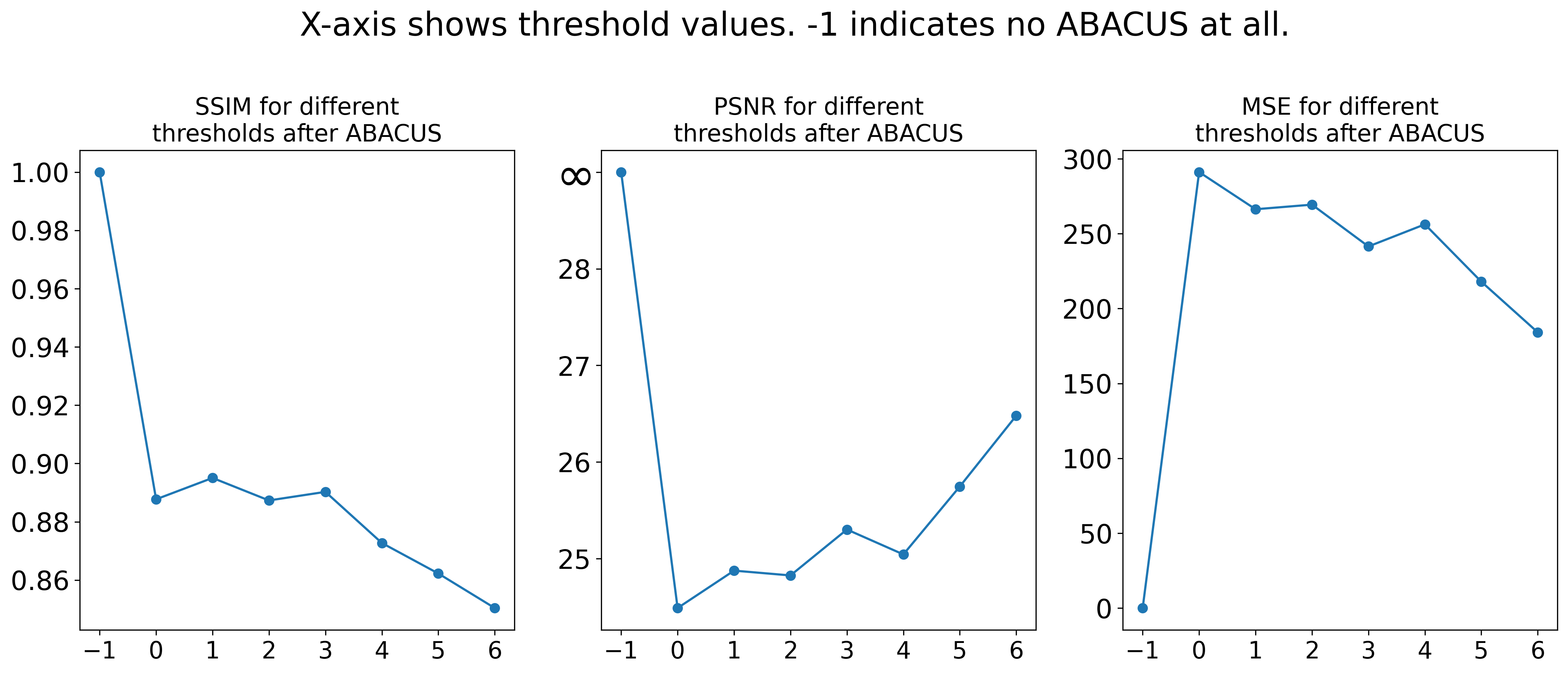}
    \caption{Image quality drop measured using three different ways. SSIM measures the total structural difference between two images and captures better how the human eye understands th difference of two images. PSNR and MSE are standard methods for image distance using signal to noise ratio and absolute difference of image pixels respectively. Note that for SSIM and PSNR the larger the value the closer are the images, while for MSE the smaller the value the closer they are.}\label{fg:psnr_ssim_mse}
\end{figure*}

We now examine the ABACUS algorithm and its application in a DTFE-based way. ABACUS gives the ability to cluster the number of initial points by globing them and moving their representative accordingly. The final points after the application of the algorithm  are fewer than the original and are weighted with a value indicating the number of original points that are globed on them. Using the intuition that galaxies located close enough may be globed to a new point to represent them and so regions dense in galaxies will end up having points with higher weights than empty regions, we made an interpolation of the ABACUS final points based on that weight. The result, shown in  Figure \ref{fg:ABACUS_interpolation}, demonstrates  some similarity with the classical DTFE   while having a smaller complexity of $O (n\log n) $.

\subsubsection{ABACUS results for down-sampling before DTFE}

In this section we demonstrate how  ABACUS can be combined with the DTFE method to obtain similar results to the original method but with significantly less computing time. The intuition behind this is that
we can exploit the  ability to end up with fewer points than the input points and our filtering (on the weight), to show that ABACUS is able to act as a pre-filter method for the application of DTFE.

This approach with only a small movement of  the original galactic positions of the initial points, results to a significantly smaller set of points which can be used as input to the DTFE algorithm, allowing its much faster completion.
Additionally,  keeping galaxies on void regions and conducting DTFE only on them, might enlighten the void-cosmological structure which is a new aspect of the problem.

As  shown in Figure \ref{fg:ABACUS_DTFE},  performing ABACUS reduces the number of galaxies to $25\%$ of the original number thus resulting in a much faster computation of the DTFE while also keeping the structure of the cosmic web. On the second row of the figure,  we further keep an even smaller number of galaxies than the original (based on the final weight). In fact on the second row right image, we observe that by keeping only $10\%$ of the original galaxies and then performing DTFE (on the reduced size  input), the structure appears to be similar with the one obtained with DTFE on the original number of galaxies. To further verify the observed results on the figures we draw insights from the field of computer science and in general the field of computer vision. We performed experiments where we used ABACUS and then filtering on the resulting points. As a final step, we applied DTFE to the filtered set of points. To compare the results, we utilize three image comparison measurements: Structural Similarity (SSIM) \cite{SSIM}, Peak signal-to-noise ratio (PSNR) \cite{PSNR}, Mean Squared Error (MSE). As we can see in Figure \ref{fg:psnr_ssim_mse}, the reduction of the input points leads to quality reduction of the resulting image as well, but this drop is rather small, 
and could be regarded as acceptable, if we take into account the computational speed up it provides. Sacrificing some of the final image quality in order to reduce the cardinality of the set points given to DTFE leads to similar results as the ones from applying DTFE to the original point set.
 We conclude that   by applying ABACUS before applying DTFE, we  manage to   reduce   the number of points, achieving a significantly better  time completion of the DTFE algorithm, while keeping the quality of the result very similar to the original one.

\subsubsection{Verification of results over time}
\begin{figure*}
    \centering
    \includegraphics[scale=.45]{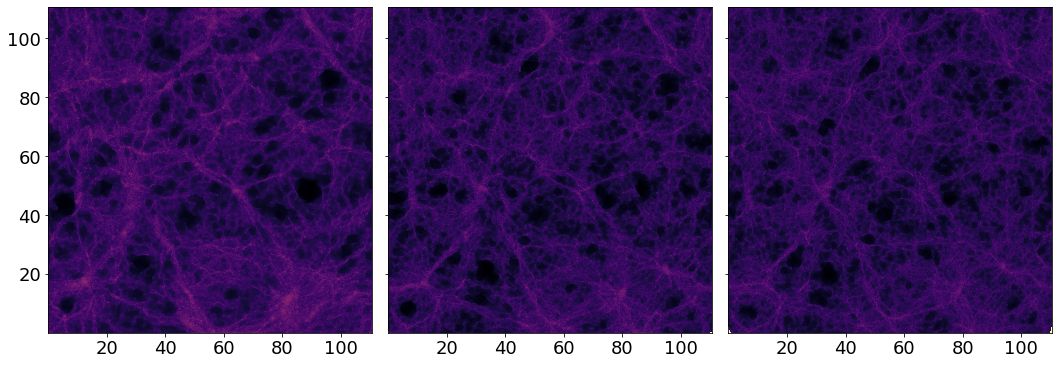}
    \caption{Time evolution of the cosmic web using the results from the DTFE method for three different redshifts (Left: $z = 0$, Middle: $z = 1$, Right: $z = 2$). As expected, for larger redshifts there are more sparse galactic structures while for redshifts closer to 0 there are more dense areas with greater structures. These results are as expected based on the effects of gravity with the passage of time and the evolution of the cosmic web. In these figures denser areas are shown with more red colors while empty or sparse areas are depicted with colors closer to black.}
    \label{fg:DTFE_time_evo}
\end{figure*}

\begin{figure*}\hphantom{.}
    \centering
    \includegraphics[scale=.4]{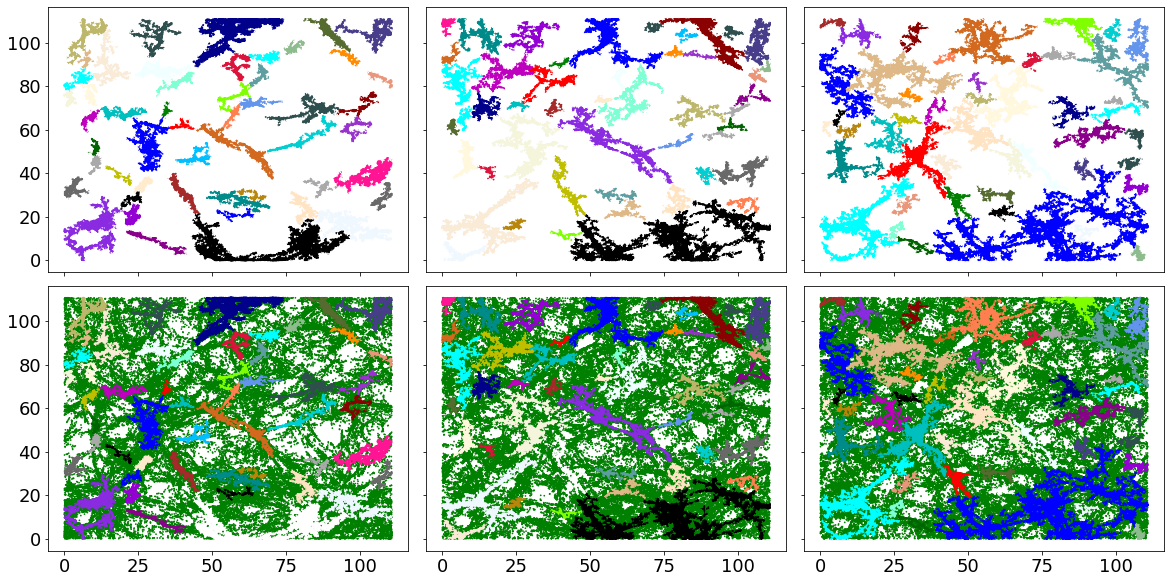}
    \caption{In this Figure we show the detected noise (void regions) in white on the top row and in green on the bottom row. On the left column we show the detected communities for $z = 0$, in the middle for $z = 1$ and on the right for $z = 2$. HDBSCAN manages to detect different communities for different redshifts following the time evolution of the cosmic web. Transforming from detecting large communities to smaller ones as time passes because as time passes dense regions tend to become denser because of the effects of gravity and thus the algorithm detects more specific communities. For example the initially blue community (down right) for $z = 2$ ends up to the white community (down left) for z=0 showing that the once large detected community ends up to a smaller yet more dense community.}
    \label{fg:HDBSCAN_time_evo}
\end{figure*}

\begin{figure*}
    \centering
    \includegraphics[scale=.5]{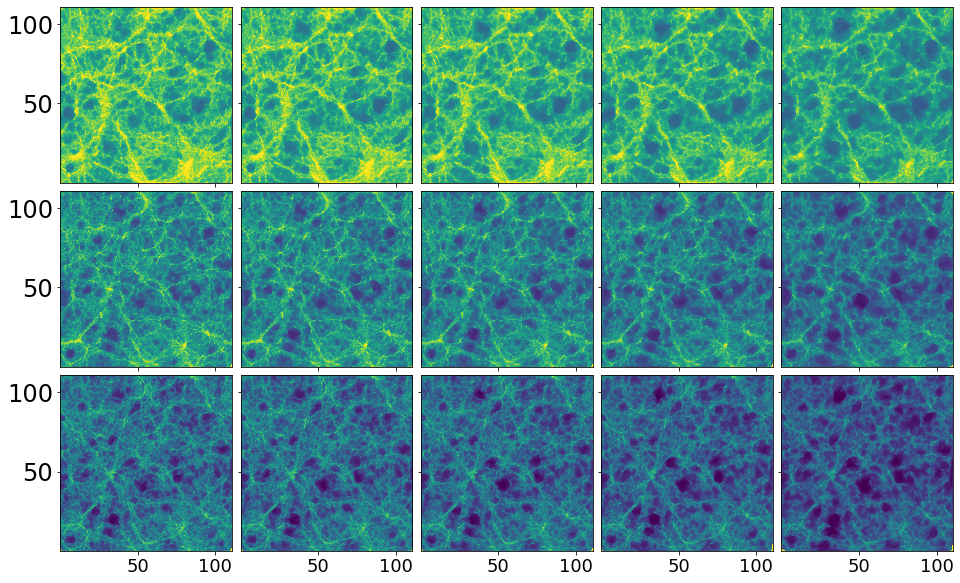}
    \caption{On the top row we show results for z = 0 and applying ABACUS with different thresholds as in Figure \ref{fg:ABACUS_DTFE} in the middle row for z = 1 and on the bottom row for z = 2. The first column shows the results from the original DTFE on the dataset while the rest of the columns show the results of applying DTFE on ABACUS results after filtering with thresholds: 0, 1, 2, 3, 4 respectively. We show that as the threshold increases the points given to DTFE decrease but as we can see we manage to give the needed points because the results of DTFE are of similar quality compared to the original DTFE. We see that without threshold (i.e. threshold value of zero) ABACUS manages to filter appropriately the data and the first two columns look similar but the second is constructed with much fewer points than the first.}
    \label{fg:ABACUS_DTFE_time_evo}
\end{figure*}

In this section we analyze briefly and solidify our results in the time evolution of the cosmic web. As we can see from Figure \ref{fg:DTFE_time_evo} the time evolution of the cosmic web follows the law of gravity. As time passes more dense and bigger galactic structures form. We can see that for large redshifts i.e. z = 2 the cosmic web is sparse and has small formation of galactic structures while by z = 0   large galactic structures with brighter colors which depict denser regions of galaxies are placed together in a more compact manner. Moreover, walls which connect clusters and voids are bigger and easier to be spotted than at z = 2. Based on this intuition, we further proceed to verify our results using HDBSCAN and ABACUS filtering before DTFE using data from different redshifts. Through this approach we verify the strength of the proposed method and observe that their success is independent of time (redshift). In Figure \ref{fg:HDBSCAN_time_evo} we show the results of HDBSCAN for three different redshifts verifying that the algorithm has detected communities change through time and manages to capture the time evolution of the cosmic web. Greater communities detected at z = 2 tear apart to smaller ones as we approach z = 0 because as time passes dense areas tend to become denser and thus HDBSCAN focuses on these areas and detects more detail in shape and size communities. Finally, in Figure \ref{fg:ABACUS_DTFE_time_evo} we show the effect of ABACUS filtering as a pre-process of the DTFE algorithm and show once again that in the time evolution application of DTFE, ABACUS can once again play a crucial role resulting in more efficient computation yet with results similar in quality as the original DTFE shown in Figure \ref{fg:DTFE_time_evo}. As we can see, the first two or even the first three columns look very similar. However,  the second and third columns are constructed using fewer points than the first column. This suggests that ABACUS can be used as a pre-processing method and can maintain good results but accelerate the computational process.

\section{Discussion and Conclusions}

Recent observational data obtained from
redshift surveys of galaxies \citet{Guzzo_2014} combined  with high quality $\Lambda$CDM simulations, provide a new era for the understanding
of the topological structure and connectivity of the cosmic web \citep{vandeWeygaert2008,Pranav_2019,Feldbrugge_2019,Wilding_2021}.


In this paper we have examined three spatial computational methods for detecting and characterizing (various parts of)  
the cosmic web such as  voids, walls, clusters and superclusters. The methods are able to reveal the structure of the cosmic web in various 
resolutions and using a variety of physical properties, such as (internal) density, size, distance and  mass. 
The methods appear to have significant results in detecting and categorizing the cosmological structures.
Their main advantage is that they can be accomplished in much faster time, making them suitable for use for larger databases of galaxies. 
This nice property allows one of them to be combined with the DTFE method to get results of similar quality but with one order of magnitude less completion time.  
More analytically:
 \begin{enumerate}
 \item We have introduced a new spatial method, called Gravity Lattice, which allows a detection of the cosmic web and could also be used in
characterization of various parts of it as  various kinds of cosmological structures, i.e. voids, walls, clusters and superclusters. 
In particular, we have implemented a 3D gravitational lattice and then we have measured the effect of the gravitational forces of the data points on nearby  
test loads of the Gravity Lattice. Furthermore,  filtering based on the number of galaxies affecting the test loads enables 
also the characterization of various parts of the cosmic web detected as  voids, walls, clusters and superclusters. 
We have assumed that the more dense (in number of galaxies) a region is, the greater is the number of galaxies affecting nearby test loads. We only show preliminary results of this method and leave further investigation as future work.

\item Inspired by a  spatial clustering algorithm called ABACUS  \citet{ABACUS}, which clusters a set of spatial points through finding a \emph{backbone} of   the data points, we have detected the backbone of the cosmic web. Furthermore, suitable modifications of ABACUS 
and appropriate filtering on the detected cosmic structures allow the detection of cosmic structures of various 
sizes and masses. 
The extraction of the backbone structure of the cosmic web can be useful for the 
understanding of the evolution of the cosmic web but can also be used as a pre-processing step 
for the application of other related algorithms from the domain, such as the DTFE algorithm, for reducing the size of the data set 
resulting to much faster methods.

\item
     We have used a  modified version of the 
HDBSCAN method  with suitable fine tuning of the values of its parameters to detect the cosmic web,  highly dense structures within it (communities) and categorizing various parts of it as voids, walls, clusters and superclusters using the over-density factor. Additionally, the method allows a hierarchical detection of high density structures.
     Also,  varying the value of the   parameter   allows the  detection of  communities of various scales, that is, communities partitioning the  
whole data set as well as communities within the detected communities.

 \end{enumerate}

 To evaluate and compare the methods explored we have utilized and combined them together with a well known standard method, namely 
the DTFE method. In particular,
 \begin{itemize}
     \item We have shown the results obtained by the HDBSCAN algorithm, indicating advantages and disadvantages of each. We have 
found that HDBSCAN outperforms in terms of time required and in that it also detects groups (communities) of highly dense cosmic structures and in that it also distinguishes high from low density structures.

     \item We have compared the results obtained by DTFE with ABACUS, showing that interpolation of the result of ABACUS can simulate up to some level the different density levels of cosmic structures detected by DTFE, illustrated with different levels of brightness.

     \item We have shown that  ABACUS can  be utilized as a pre-processing step of the  DTFE method. The resulting algorithm maintains the advantages of both algorithms: it is fast and the detected cosmic web is almost the same as with the original DTFE.

     \item We have used data from different redshifts and applied the above methods on them in order to study briefly the evolution 
of the cosmic web through these new methods we have presented. Through this approach we manage to verify that they 
are able to capture different formations of the cosmic web as they appear in time while maintaining low completion time.
 \end{itemize}

To summarize in our view the advantages of the explored methods are multiple. 
On one hand they have the ability to be accomplished in significantly less time than the DTFE method. 
Furthermore we have explored the ability of ABACUS to be combined with DTFE, to get similar results in significantly less time. 
All these make them a powerful computational tool for the study and understanding of the large datasets of real and simulated cosmological data that are currently available.

\section*{Acknowledgements}

The authors acknowledge support from the project EXCELLENCE/1216/0207/ GRATOS funded by the Cyprus Research \& Innovation Foundation.

\section*{Data Availability Statement}

The data underlying this article are available at the public database of the IllustrisTNG project.



 \bibliographystyle{mnras}
 \bibliography{biblio}








\bsp	
\label{lastpage}
\end{document}